\documentclass[a4paper,fleqn,usenatbib]{mnras}

\usepackage{newtxtext,newtxmath}

\usepackage[T1]{fontenc}
\usepackage{ae,aecompl}

\usepackage{graphicx}	
\usepackage{amsmath}	
\usepackage{amssymb}	

\newcommand{\be}{\begin{equation}}
\newcommand{\ee}{\end{equation}}
\newcommand{\ba}{\begin{eqnarray}}
\newcommand{\ea}{\end{eqnarray}}
\newcommand{\lp}{\left(}
\newcommand{\rp}{\right)}
\newcommand{\lb}{\left[}
\newcommand{\rb}{\right]}
\newcommand{\Msun}{M_\odot}
\newcommand{\rlc}{r_{\rm lc}}
\newcommand{\rmag}{r_{\rm m}}
\newcommand{\rco}{r_{\rm co}}
\newcommand{\Omegas}{\Omega}
\newcommand{\OmegaK}{\Omega_{\rm K}}
\newcommand{\PK}{P_{\rm eq}}
\newcommand{\tem}{t_{\rm em}}
\newcommand{\tej}{t_{\rm ej}}
\newcommand{\tprop}{t_{\rm prop}}
\newcommand{\mcsnr}{XMMU J051342.6$-$672412}

\title[Early NS evolution in HMXBs]{Early neutron star evolution in high-mass X-ray binaries}

\author[W. C. G. Ho et al.]{Wynn C. G. Ho$^{1,2}$\thanks{E-mail: wynnho@slac.stanford.edu},
M. J. P. Wijngaarden,$^{2}$
Nils Andersson,$^{2}$
Thomas M. Tauris$^{3,4}$
\newauthor
and F. Haberl$^{5}$
\\
$^{1}$Department of Physics and Astronomy, Haverford College, 370 Lancaster Avenue, Haverford, PA, 19041, USA\\
$^{2}$Mathematical Sciences and STAG Research Centre, University of Southampton, SO17 1BJ, Southampton, UK\\
$^{3}$Aarhus Institute of Advanced Studies (AIAS), Aarhus University, H{\o}egh-Guldbergs~Gade~6B, 8000~Aarhus~C, Denmark\\
$^{4}$Department of Physics and Astronomy, Aarhus University, Ny Munkegade 120, 8000~Aarhus~C, Denmark \\
$^{5}$Max-Planck-Institut f\"ur extraterrestrische Physik, Giessenbachstra{\ss}e, 85748 Garching, Germany
}

\date{Accepted 2020 March 5. Received 2020 March 3; in original form 2020 January 12}

\pubyear{2020}

\begin{document}
\label{firstpage}
\pagerange{\pageref{firstpage}--\pageref{lastpage}}
\maketitle

\begin{abstract}
The application of standard accretion theory to observations of
X-ray binaries provides valuable insights into neutron star
properties, such as their spin period and magnetic field.
However, most studies concentrate on relatively old systems,
where the neutron star is in its late propeller, accretor, or
nearly spin equilibrium phase.
Here we use an analytic model from standard accretion theory to
illustrate the evolution of high-mass X-ray binaries early in their life.
We show that a young neutron star is unlikely to be an accretor
because of the long duration of ejector and propeller phases.
We apply the model to the recently discovered $\sim 4000\mbox{ yr}$
old high-mass X-ray binary \mcsnr\ and find that the system's
neutron star, with a tentative spin period of 4.4~s, cannot be in
the accretor phase and has a magnetic field
$B>\mbox{a few}\times 10^{13}\mbox{ G}$, which is
comparable to the magnetic field of many older high-mass X-ray binaries and is
much higher than the spin equilibrium inferred value of
$\mbox{a few}\times 10^{11}\mbox{ G}$.
The observed X-ray luminosity could be the result
of thermal emission from a young cooling magnetic neutron star
or a small amount of accretion that can occur in the propeller phase.
\end{abstract}




\section{Introduction} \label{sec:intro}

High-mass X-ray binaries (HMXBs) have typical ages of
$\sim 10^6-10^7\mbox{ yr}$, based on the main sequence and post-main
sequence lifetimes of the high-mass companion star.
However, for a few HMXBs, an association with a supernova remnant
has been made, which limits their age to $< 10^5\mbox{ yr}$
\citep{haberletal12,henaultbrunetetal12,sewardetal12,heinzetal13,gvaramadzeetal19,maitraetal19}.
Possibly the youngest HMXB is the one recently discovered near the
geometrical centre of the supernova remnant MCSNR~J0513$-$6724,
with an age of $\approx 3800_{-900}^{+1900}\mbox{ yr}$.
This HMXB, which we name \mcsnr\ based on the coordinates derived
using {\it XMM-Newton} data,
has X-ray luminosity $\sim 7\times 10^{33}\mbox{ erg s$^{-1}$}$
and a likely neutron star (NS) component with a spin period of
$4.4\mbox{ s}$ \citep{maitraetal19}.
Detections of young HMXBs provide snapshots early in the
accretion history and spin evolution of NS/pulsars and can yield
valuable insights into a hitherto unknown stage of HMXB evolution
and accretion theory.

\begin{figure}
\includegraphics[width=\columnwidth]{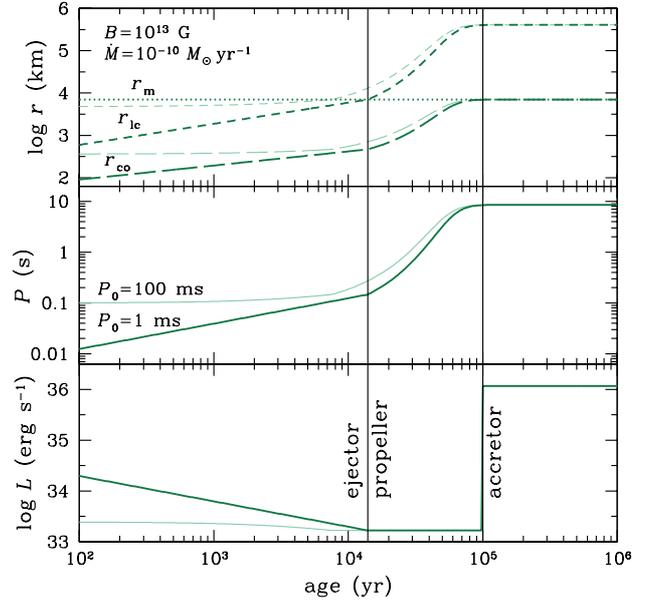}
\caption{
Top panel: Evolution of light cylinder radius $\rlc$ (short-dashed),
magnetosphere radius $\rmag$ (dotted), and corotation radius $\rco$
(long-dashed) for constant magnetic field $B=10^{13}\mbox{ G}$,
constant accretion rate $\dot{M}=10^{-10}\,\Msun\mbox{ yr$^{-1}$}$, and
initial spin periods $P=1$ (dark) and 100~ms (light).
Vertical lines separate ejector ($\rmag>\rlc$), propeller ($\rco<\rmag<\rlc$),
and accretor/spin-equilibrium phases ($\rmag\lesssim\rco$).
Middle panel: Spin period evolution as determined by
equations~(\ref{eq:evolej}) and (\ref{eq:evolprop}).
Bottom panel: Luminosity evolution as determined by equation~(\ref{eq:lum}).
}
\label{fig:phases}
\end{figure}

Most NSs are inferred to be born with a magnetic field
$B\sim 10^{13}\mbox{ G}$ and spin period $P\sim 100\mbox{ ms}$
(e.g., \citealt{fauchergiguerekaspi06,gullonetal14}).
For example, the $\sim$1000~yr old Crab Pulsar has
$B=4\times 10^{12}\mbox{ G}$ and $P=33\mbox{ ms}$.
Traditional NS accretion theory dictates that this spin period is too
short to allow matter inflowing at a rate $\dot{M}$ to accrete onto
the NS because it cannot penetrate the pulsar light cylinder, which
is at distance
\be
\rlc = c/\Omega = 48\mbox{ km}\lp P/1\mbox{ ms}\rp, \label{eq:rlc}
\ee
where $\Omegas\equiv 2\pi/P$
\citep{shvartsman71,illarionovsunyaev75,lipunov92}.
In other words, for spin periods
\be
P < 2\pi\rmag/c = 150\mbox{ ms }B_{13}^{4/7}\dot{M}_{-10}^{-2/7},
\label{eq:pej}
\ee
$\rlc$ is smaller than the size of the magnetosphere, and the
NS is in the {\it ejector} phase.
Here we adopt the conventional approximation for magnetosphere size
(e.g., \citealt{pringlerees72,lambetal73,davidsonostriker73})
\be
\rmag = \xi r_{\rm A} = \xi\lp\frac{\mu^4}{8GM\dot{M}^2}\rp^{1/7}
 = 7.0\times 10^3\mbox{ km }B_{13}^{4/7}\dot{M}_{-10}^{-2/7},
 \label{eq:rmag}
\ee
where $\xi\approx 0.5$
(e.g., \citealt{ghoshlamb79,wang96,campanaetal18,chashkinaetal19,vasilopoulosetal20}),
the Alfv\'en radius $r_{\rm A}$ is derived from balancing the ram pressure
of accreting matter with pressure of the pulsar magnetic field,
$\mu=BR^3/2$ is the magnetic dipole
moment\footnote{Note the factors of 8 and 2 in the denominator of
$\rmag$ and $\mu$, respectively, in contrast to other definitions
in the literature which have different factors of order unity;
this implies that derived values of various parameters such as $B$ can
differ by a factor of a few if these alternative definitions are used.},
$B_{13}=B/10^{13}\mbox{ G}$,
$\dot{M}_{-10}=\dot{M}/10^{-10}\,\Msun\mbox{ yr$^{-1}$}$, and we assume
a NS mass $M=1.4\,\Msun$ and radius $R=10\mbox{ km}$.
Once the NS slows down sufficiently by electromagnetic dipole radiation,
matter enters the light cylinder but is still unable to accrete onto
the NS surface due to the centrifugal barrier.
Instead, the pulsar is spun down by the torque of matter being flung
out when $\rmag$ is greater than the corotation radius
\be
\rco = \lp GM/\Omega^2\rp^{1/3}
 = 17\mbox{ km}\lp P/1\mbox{ ms}\rp^{2/3}, \label{eq:rco}
\ee
and the pulsar is in the {\it propeller} phase.
Once $\rmag\lesssim\rco$, the pulsar can be spun-up by gaining the angular
momentum carried by infalling matter in the {\it accretor} phase.
We note that, when $\rmag\sim\rco$ (with the precise values being
uncertain, depending on the critical fastness parameter $\hat{\omega}_{\rm s}$;
\citealt{elsnerlamb77,ghoshlamb79,wang95}), spin-down and spin-up torques
balance such that the net torque on the pulsar is nearly zero, the
spin period does not change, and the pulsar is in {\it spin equilibrium}
\citep{davidsonostriker73,illarionovsunyaev75}. 
The fastness parameter $\hat{\omega}_{\rm s}\equiv\Omegas/\OmegaK(\rmag)$,
where the Keplerian orbital frequency $\OmegaK(\rmag)$ at the
magnetosphere radius has the corresponding (spin equilibrium) period
\be
\PK=\frac{2\pi}{\OmegaK(\rmag)} = \lp\frac{4\pi^2\rmag^3}{GM}\rp^{1/2}
 = 8.5\mbox{ s }B_{13}^{6/7}\dot{M}_{-10}^{-3/7}. \label{eq:peq}
\ee
In real systems, small variations in accretion rate can cause
small spin period time derivatives $\dot{P}$ and deviations from
spin equilibrium.
This may result in accretion onto the NS surface, as seen in
observations, as well as in numerical simulations which model
three-dimensional structure, viscosity, mass loss, and other effects (e.g.,
\citealt{lovelaceetal95,romanovaetal04,shakuraetal12,taurisetal12,shietal15,parfreyetal17}).
The characteristic lengthscales ($\rlc$, $\rmag$, and $\rco$) are plotted
for $B=10^{13}\mbox{ G}$ and $\dot{M}=10^{-10}\,\Msun\mbox{ yr$^{-1}$}$
in Figure~\ref{fig:phases}, where we also highlight the ejector,
propeller, and accretor/spin-equilibrium phases implied by the
relative values of $\rlc$, $\rmag$, and $\rco$.
For systems evolving with mass-transfer via Roche-lobe overflow, 
spin equilibrium is disrupted again at the Roche-lobe decoupling phase \citep{tau12}.

In this work, we use the simple analytic spin period evolution model
of \citet{hoandersson17}, which is based on standard accretion theory
(see, e.g., \citealt{ghoshlamb79,wang87,lipunov92}),
to illustrate the evolution of the NS in HMXBs and expected
accretion state of these systems, especially young ones like \mcsnr.
Section~\ref{sec:model} describes the evolution model.
Section~\ref{sec:mcsnr} applies the model to \mcsnr.
Section~\ref{sec:discuss} summarizes our work and discusses implications and some of the model assumptions.

\section{Model for spin evolution and accretion phases} \label{sec:model}

As mentioned in Section~\ref{sec:intro}, a pulsar in the ejector
phase does not interact with accreting matter and spins down as
if in isolation, i.e., by emission of dipole radiation, such that $d\Omegas/dt=-\beta\Omegas^3$, where
\be
\beta\equiv2\mu^2/3c^3I=B^2R^6/6c^3I=6.2\times10^{-16}\mbox{s }B_{13}^2
\ee
and we assume a NS moment of inertia $I=10^{45}\mbox{ g cm$^{2}$}$.
For simplicity, we use the traditional vacuum dipole formula of
\citet{pacini68,gunnostriker69} and consider an orthogonal rotator.
Thus the spin evolution from an initial spin rate $\Omegas_0$
($=2\pi/P_0$) is
\be
\Omega = \Omega_0\lp 1+2\beta\Omega_0^2t\rp^{-1/2}\!\!\!\!
 = \Omega_0\lp 1+t/\tem\rp^{-1/2} \quad\mbox{for $t<\tej$},
\label{eq:evolej}
\ee
where
\be
\tem=1/\lp 2\beta\Omega_0^2\rp=0.65\mbox{ yr }B_{13}^{-2}\lp P_0/1\mbox{ ms}\rp.
\label{eq:tem}
\ee
The ejector phase lasts until $t=\tej$ when $\rmag=\rlc$, where
equations~(\ref{eq:pej}) and (\ref{eq:evolej}) give
\be
\tej = \tem\lb\lp\frac{\Omega_0\rmag}{c}\rp^2-1\rb
 \approx \frac{\rmag^2}{2\beta c^2}
 = 1.4\times 10^4\mbox{ yr }B_{13}^{-6/7}\dot{M}_{-10}^{-4/7}.
\label{eq:tej}
\ee

Once the propeller phase begins, the spin evolution is governed approximately
by (e.g., \citealt{illarionovsunyaev75,alpar01,hoetal14,hoandersson17};
see also \citealt{parfreyetal16})
\be
I\frac{d\Omega}{dt} = -\dot{M}\rmag^2\lb\Omega-\OmegaK(\rmag)\rb
 = \frac{I\OmegaK}{\tprop}\lp 1-\hat{\omega}_{\rm s}\rp,
\label{eq:evolpropeq}
\ee
where
\be
\tprop \equiv I/\dot{M}\rmag^2
 = 1.0\times 10^4\mbox{ yr }B_{13}^{-8/7}\dot{M}_{-10}^{-3/7}.
\label{eq:tprop}
\ee
One term is the propeller/spin-down torque, while the other
term is the accretion/spin-up torque.
A simple solution of equation~(\ref{eq:evolpropeq}) can be obtained
by assuming constant $\mu$ and $\dot{M}$ (and thus constant $\rmag$ and $\OmegaK$), yielding
\be
\Omega = \lb\Omega_{\rm ej}-\OmegaK(\rmag)\rb e^{-(t-\tej)/\tprop}
 +\OmegaK(\rmag)\quad\mbox{for $t>\tej$},
\label{eq:evolprop}
\ee
where $\Omega_{\rm ej}\equiv c/\rmag$ is the spin frequency corresponding
to the critical spin period marking the end of the ejector phase and
beginning of the propeller phase [see equation~(\ref{eq:pej})].
One can see from equation~(\ref{eq:evolprop}) that, once the spin
rate evolves to the point when $\rco=\rmag$, the term in brackets
cancel and the spin period is constant at the spin equilibrium value
$\PK$ given by equation~(\ref{eq:peq}).

Equations~(\ref{eq:evolej}) and (\ref{eq:evolprop}) describe the
complete evolution of NS spin frequency (or spin period) from the
ejector phase, through to the propeller phase, and then to the
accretor/spin equilibrium phase.
The middle panel of Figure~\ref{fig:phases} shows this evolution for a
NS with $B=10^{13}\mbox{ G}$, $\dot{M}=10^{-10}\Msun\mbox{ yr$^{-1}$}$
and initial spin periods $P_0=1$ and 100~ms.
As is clear, the choice of initial spin period makes no difference to
the evolution of the spin period at later times, such as in the
accretor phase.

We can obtain an estimate of the HMXB luminosity during the different
accretion phases due simply to gravitational infall
\be
L = GM\dot{M}/r = 1.2\times 10^{36}\mbox{ erg s$^{-1}$}
\dot{M}_{-10}\lp10\mbox{ km}/r\rp, \label{eq:lum}
\ee
where $r=\rlc$ during the ejector phase, $r=\rmag$ during the
propeller phase, and $r=R$ during the accretor/spin equilibrium phase,
and no beaming is assumed (e.g., \citealt{kingcominsky94,stellaetal94}).
Two examples are shown in the bottom panel of Figure~\ref{fig:phases}.
The above estimate assumes the accretion flow is cold and does not
radiate on its own and thus represents a minimum luminosity, as
other emission processes could contribute and dominate the observed
flux from an accreting system.

\begin{figure}
\includegraphics[width=\columnwidth]{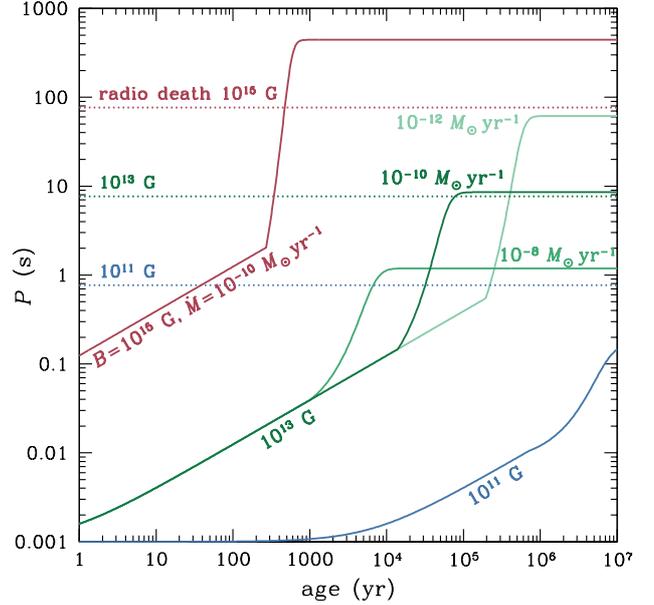}
\caption{
Spin period as a function of time, starting from ejector phase
onset at $P_0=1\mbox{ ms}$,
for magnetic fields $B=10^{11}$, $10^{13}$, and $10^{15}\mbox{ G}$ and
accretion rate $\dot{M}=10^{-10}\Msun\mbox{ yr$^{-1}$}$
and for $B=10^{13}\mbox{ G}$ and $\dot{M}=10^{-12}$
and $10^{-8}\Msun\mbox{ yr$^{-1}$}$.
Horizontal dotted lines indicate the theoretically uncertain death
line for radio pulsar emission for the magnetic fields shown.
}
\label{fig:spinage}
\end{figure}

The dependence of spin period evolution on magnetic field $B$ and
accretion rate $\dot{M}$ is illustrated in Figure~\ref{fig:spinage}.
Also plotted are radio pulsar death lines, above which radio emission
is thought to be inoperative \citep{sturrock71,rudermansutherland75}.
The exact location and dependencies of the death line are uncertain,
and we simply use $P=7.7\mbox{ s }B_{13}^{1/2}$ from
\citet{bhattacharyaetal92}; for alternative death lines, see, e.g.,
\citet{chenruderman93,zhangetal00,hibschmanarons01}.
One can see that, in some cases, even if radio emission is not suppressed
by accretion, radio emission would still be inactive.

\section{Application to \mcsnr} \label{sec:mcsnr}

In this section, we apply the simple accretion model to the
youngest NS with a known spin period in a HMXB, \mcsnr.
\citet{maitraetal19} recently identified \mcsnr\ as a HMXB at the
center of the Large Magellanic Cloud supernova remnant MCSNR~J0513$-$6724.
The size of the supernova remnant yields
an age of $3800_{-900}^{+1900}\mbox{ yr}$.
An OGLE light curve shows the B2.5Ib optical counterpart to have a 2.2~d
periodicity, which is interpreted as the orbital period.
{\it XMM-Newton} data reveal pulsations at 4.4~s, which is interpreted
as the NS spin period, and a power law spectrum with a 0.2--12~keV
luminosity of $7\times 10^{33}\mbox{ erg s$^{-1}$}$ (at 50~kpc).
\citet{maitraetal19} then attribute the measured luminosity
to matter accreting onto the NS surface
[equation~(\ref{eq:lum}) with $r=R$], which implies a mass accretion
rate of $\dot{M}=6\times 10^{-13}\Msun\mbox{ yr$^{-1}$}$,
and derive a magnetic field $B\sim 4\times 10^{11}\mbox{ G}$,
assuming the NS is at spin equilibrium [using equation~(\ref{eq:peq})].

This result for \mcsnr\ is problematic in standard accretion
theory because the described scenario does not account for evolution.
If the magnetic field is indeed as low as $\sim 4\times10^{11}\mbox{ G}$,
then the duration of the ejector phase from equation~(\ref{eq:tej}) is
$\tej=2\times 10^5\mbox{ yr }\dot{M}_{-10}^{-4/7}$,
and this would only be comparable to the age of \mcsnr\ for an accretion
rate $\dot{M}\sim 10^{-7}\Msun\mbox{ yr$^{-1}$}$, greatly exceeding the
accretion rate onto the NS surface implied by the observed luminosity.
On the other hand, with such a low field, the pulsar spin period would
not have changed significantly from its value at birth, which means that
it would have been born in the propeller phase [see equation~(\ref{eq:pej})].
In fact, the propeller phase would also be long, with a timescale
from equation~(\ref{eq:tprop}) of
$\tprop=4\times 10^5\mbox{ yr }\dot{M}_{-10}^{-3/7}$, unless the
accretion rate is an even higher
$\dot{M}\sim 5\times 10^{-6}\Msun\mbox{ yr$^{-1}$}$.
In summary, with a magnetic field as low as
$\mbox{a few}\times 10^{11}\mbox{ G}$, the NS in \mcsnr\ would not have
had enough time to slow down sufficiently to be in the accretor phase,
unless the accretion rate greatly exceeds observations and expectations.

Let us apply the analytic model of Section~\ref{sec:model}, which
qualitatively encapsulates standard accretion theory and evolution
of accreting systems, to infer the possible accretion phase,
magnetic field, and accretion rate of \mcsnr.
In doing so, we must match the observed values of spin period
$P=4.4\mbox{ s}$ and luminosity $7\times 10^{33}\mbox{ erg s$^{-1}$}$
at an age of $\approx$2900--5700~yr.
For simplicity, we do not apply a bolometric correction to the observed
X-ray luminosity.

First, we consider what criteria are needed for \mcsnr\ to be
in the ejector phase.  In this phase, energy loss from
electromagnetic dipole radiation drives spin period evolution,
which is described by equation~(\ref{eq:evolej}).
From $\mbox{age}=(P/2\pi)^2/(2\beta)$, where $P=4.4\mbox{ s}$ and
the age is 2900--5700~yr, we find that the magnetic field must be
$B=(5-7)\times 10^{14}\mbox{ G}$.
At greater fields strengths, spin-down is too effective, and \mcsnr\
would have a much longer spin period at the current age.
The accretion rate must also be low enough such that the current spin
period is below the limit needed to initiate the propeller phase.
From equation~(\ref{eq:pej}), we find that
$\dot{M}<2\times 10^{-12}\Msun\mbox{ yr$^{-1}$}$.
Finally, the light cylinder radius $\rlc=2.1\times 10^5\mbox{ km}$,
such that the accretion luminosity is
$L=6\times 10^{31}\mbox{ erg s$^{-1}$}$, which is well below
the observed luminosity.

\begin{figure}
\includegraphics[width=\columnwidth]{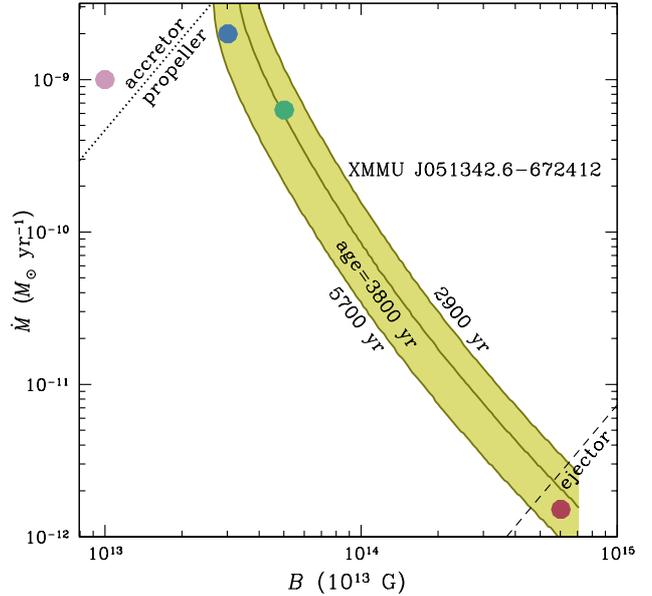}
\caption{
Constraints on the magnetic field $B$ and accretion rate
$\dot{M}$ of \mcsnr.
The shaded region denotes $B$ and $\dot{M}$ values which produce
a NS with the observed 4.4~s spin period at the 2900--5800~yr age
of \mcsnr.
Colored filled circles denote combinations of $B$ and $\dot{M}$ whose
spin period and luminosity evolutions are shown in Figure~\ref{fig:j0513}.
Dotted line and short-dashed line separate regions where the NS is in
accretor, propeller, and ejector phases.
}
\label{fig:j0513_magbmdot}
\end{figure}

From the above discussion, we expect that
if the magnetic field is below that of the ejector phase and
accretion rate is higher, then the NS will be in the propeller state.
The shaded region in Figure~\ref{fig:j0513_magbmdot}
illustrates the relation between $B$ and $\dot{M}$ needed to solve
the evolution given by equation~(\ref{eq:evolprop}), i.e.,
\be
\ln\frac{\Omega_{\rm ej}-\OmegaK}{\Omega-\OmegaK}
 =\frac{|\mbox{age}-\tej|}{\tprop},
\ee
for \mcsnr.
Regions where values of $B$-$\dot{M}$ would produce an accretor and
ejector are also indicated in Figure~\ref{fig:j0513_magbmdot}.

\begin{figure}
\includegraphics[width=\columnwidth]{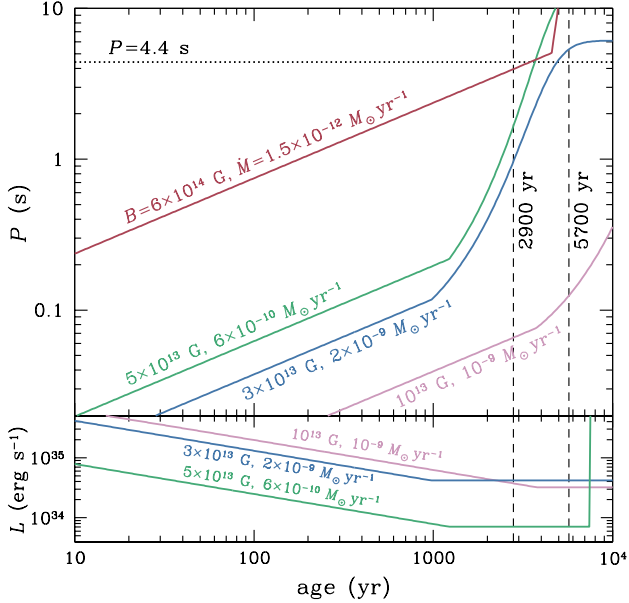}
\caption{
Top panel: Spin period evolutions for magnetic field and
accretion rate [$B$(G),$\dot{M}$($\Msun\mbox{ yr$^{-1}$}$)]
=[$10^{13}$, $10^{-9}$], [$3\times 10^{13}$, $2\times 10^{-9}$],
[$5\times 10^{13}$, $6\times 10^{-10}$], and
[$6\times 10^{14}$, $1.5\times 10^{-12}$] and $P_0=1\mbox{ ms}$.
Horizontal dotted line indicates the current 4.4~s spin period of \mcsnr,
and vertical dashed lines denote bounds on its age.
Bottom panel: Luminosity evolutions corresponding to the spin period
evolutions shown in the top panel, where luminosity is calculated
using equation~(\ref{eq:lum}) and $r=\rlc$ during the ejector phase,
$r=\rmag$ during the propeller phase, and $r=R$ during the
accretor/spin equilibrium phase.
The luminosity evolution for [$6\times 10^{14}$, $1.5\times 10^{-12}$]
is below the luminosity range displayed.
}
\label{fig:j0513}
\end{figure}

Figure~\ref{fig:j0513} shows evolutions of spin period (upper panel)
and luminosity (lower panel) for various combinations of magnetic
field $B$ and accretion rate $\dot{M}$ that lead to ejector and
propeller phases for \mcsnr\ (see Figure~\ref{fig:j0513_magbmdot}).
Note that the relatively low $B=10^{13}\mbox{ G}$ evolution is for
the accretor phase but only at an age much older than that of \mcsnr.
The evolution of this case
produces a luminosity which exceeds that seen from \mcsnr.

The origin of \mcsnr's observed pulsed X-ray luminosity
($L_{\rm X}\sim 7\times 10^{33}\mbox{ erg s$^{-1}$}$) is uncertain.
If the NS is in the propeller phase, then the accretion luminosity
[equation~(\ref{eq:lum})] at the magnetosphere would match the
observed X-ray luminosity for $B\approx(3-6)\times 10^{13}\mbox{ G}$,
although the temperature at the relevant $\rmag$ ($\sim 10^4\mbox{ km}$)
is probably too low to result in much X-ray emission.
But even in the propeller phase, a small amount of matter can reach the
NS surface intermittently
\citep{romanovaetal04,romanovaetal18,dangelospruit10,ds12}
since the centrifugal barrier only applies to the closed portion of
the magnetosphere, and the X-ray luminosity of \mcsnr\ would imply a
low residual accretion rate $\sim 6\times 10^{-13}\Msun\mbox{ yr$^{-1}$}$,
which is comparable to (uncertain) theoretical estimates in the
strong ($\hat{\omega}_{\rm s}\gg 1$) propeller regime (e.g.,
\citealt{lipunovshakura76,menouetal99}; see also \citealt{gungoretal17}).
On the other hand, if the NS is in the accretor phase, then accretion
directly onto the NS surface at
$\dot{M}>3\times 10^{-9}\Msun\mbox{ yr$^{-1}$}$
(see Figure~\ref{fig:j0513_magbmdot})
would produce $L_{\rm X}>4\times 10^{37}\mbox{ erg s$^{-1}$}$.
Another possibility is that \mcsnr\ emits like young rotation-powered
pulsars, which can produce pulsed non-thermal X-rays from their
magnetosphere and pulsar wind nebula with X-ray luminosities
$<10^{-2}\dot{E}$, where $\dot{E}=I\Omegas\dot{\Omegas}$ is
spin-down power \citep{beckertruemper97,enotoetal19}.
The two intermediate cases shown in Figure~\ref{fig:j0513} have
too low $\dot{E}$ ($\sim 10^{35}\mbox{ erg s$^{-1}$}$), but it is
important to remember that the $L_{\rm X}$--$\dot{E}$ correlation
holds for rotation-powered pulsars.
Finally, we find that a 0.41~keV blackbody model can fit the spectra
of \mcsnr\ slightly better than the power law model fit of
\citet{maitraetal19} (blackbody C-statistic $=117.3$ versus
power law C-statistic $=124.6$ for 126 degrees of freedom).  The
resulting 1~km emission radius could indicate thermal radiation from
a hot spot on the NS surface, and a strong surface magnetic field can
contribute to generating the observed strong pulsations
(see, e.g., PSR J1119$-$6127; \citealt{ngetal12}).
Thus the bulk of X-rays from \mcsnr\ could be thermal emission from
a young X-ray dim isolated NS (XDINS) or magnetar since both share
some similar properties, e.g., XDINSs and magnetars have
$P\approx 2-17\mbox{ s}$, $B>10^{13}\mbox{ G}$, and
quiescent X-ray luminosity $\sim 10^{31}-10^{36}\mbox{ erg s$^{-1}$}$
\citep{haberl07,turolla09,kaspibeloborodov17,enotoetal19,huetal19}.

\section{Discussion} \label{sec:discuss}

In this work, we considered a simple analytic model that follows from
standard accretion theory (see also \citealt{hoandersson17}).
We showed how the NS spin period evolves during ejector and propeller
phases, until reaching accretor and spin equilibrium phases, and
provide estimates of the ejector phase duration and propeller phase
timescale.
We also estimated the evolution of accretion luminosity, which scales as
$1/r$, where $r$ can be equal to the light cylinder radius $\rlc$,
magnetosphere radius $\rmag$, or NS radius $R$ during ejector,
propeller, or accretor/spin equilibrium phases, respectively.
Applying this model to the recently discovered young HMXB \mcsnr,
we inferred that the NS is likely to be in the propeller phase
and that the NS has a magnetic field $B>\mbox{a few}\times 10^{13}\mbox{ G}$.
This magnetic field is stronger than, but comparable to,
the magnetic field measured or inferred in many other HMXBs
(e.g., \citealt{hoetal14,staubertetal19}).
The observed X-ray luminosity could be due to
thermal emission from the magnetised surface of this young cooling NS
or a small amount of matter that leaks through the centrifugal barrier
and accretes onto the NS surface.

While the simple model is based on fundamentals of
standard accretion theory, caution must be exercised in using the
precise results.
Real accreting systems are complex and require detailed modeling and
numerical simulations for accurate quantitative solutions
(see, e.g.,
\citealt{lovelaceetal95,romanovaetal04,romanovaetal18,parfreyetal17}).
For example, \citet{spk+15} develop a more detailed model of wind
accretion onto a NS in HMXB systems. They distinguish between two main
regimes: Supersonic (Bondi-Hoyle-Lyttleton) accretion when captured
matter cools rapidly and falls supersonically towards the NS
magnetosphere and subsonic (settling) accretion when captured hot
plasma reaches the magnetosphere boundary.
In the first regime, shocked matter cools via Compton processes and
enters the magnetosphere due to Rayleigh-Taylor instabilities
\citep{al76}. The accreting NS can either spin up or spin down,
depending on whether the wind carries prograde or retrograde angular
momentum when entering the magnetosphere
(see also \citealt{shapirolightman76,wang81,klusetal14}).
In the second regime, matter remains hot since the plasma cooling
time is much longer than the free-fall time, and a quasi-static shell
forms around the magnetosphere leading to subsonic accretion. In this
case, both spin-up and spin-down can occur even if the specific
angular momentum of the wind is only prograde. \citet{spk+15} argue
that triggering of the transition from supersonic to subsonic accretion
may be related to a switch in the X-ray beam pattern in response
to a change in optical depth, i.e., the X-ray beam pattern changes
with decreasing X-ray luminosity
(near $4\times 10^{36}\mbox{ erg s$^{-1}$}$) from a fan beam to
a pencil beam. Observational evidence to support this hypothesis
is found in pulse profile observations of Vela~X-1 in different
energy bands \citep{dss11}.

In the case of \mcsnr, $L_X\sim 7\times 10^{33}\mbox{ erg s$^{-1}$}$
and the plasma is expected to remain hot until it reaches the
magnetosphere boundary. As a consequence, the effective gravitational
acceleration changes above the magnetosphere, and the average radial
velocity of the settling plasma is expected to be smaller than the
standard free-fall velocity ($=\sqrt{2GM/r}$). This correction would
produce some changes in our numerical results. 
Moreover, the effects of a complicated non-stationary accretion wake
\citep{ec15,dkss17} and a trapped disc with cyclic accretion
\citep{ds12} are difficult to quantify, and a full treatment of these
effects is beyond the scope of this paper.

Other caveats are our assumptions of constant accretion rate and
magnetic field throughout the various accretion phases.
Modelling of stellar evolution in binaries shows that the wind
mass-loss rate, and thus mass-transfer rate, varies significantly
over time \citep{lan12}. This is also true for mass transfer via
Roche-lobe overflow \citep{taurisetal12}. Meanwhile, the magnetic
field of an accreting NS is expected to decay, e.g., as a consequence
of heating of the crust, which reduces its electrical conductivity
\citep{romani90,bha02}.
On the other hand, a magnetic field that is increasing at the present
time could be the
result of an early episode of field burial by accretion at very high
rates (in order to prevent field re-emergence on short timescales;
\citealt{chevalier89,geppertetal99,bernaletal10,ho11,ho15,viganopons12}).
However, a re-emerging magnetic field would not allow \mcsnr\
to be in the accretor phase at the current time since the field
would have been weaker in the past.
A weak field would produce a long timescale for spin-down
[equation~(\ref{eq:tem})].
Thus \mcsnr\ would not reach the current spin period of 4.4~s if it
was born at typical birth periods of less than one second
(if \mcsnr\ was born near its current period, the propeller timescale
at these weak fields is longer than the current age;
see discussion in Section~\ref{sec:mcsnr}).
A time-varying accretion rate
(see, e.g.,
\citealt{taurisetal12,bhattacharyyachakrabarty17,dangelo17,mushtukovetal19})
might yield a solution such that
\mcsnr\ is in the accretor phase and have a weak field, but this
would require fine-tuning.  For example, a very high accretion
rate at early times could cause cessation of the ejector phase,
but this would also shorten the propeller phase such that the
accretor/spin equilibrium phase would begin at much shorter periods
than the current spin period.

Circinus~X-1 is another (possible) HMXB in a young ($<4600\mbox{ yr}$)
supernova remnant \citep{heinzetal13}.
Circinus~X-1 is identified as a NS system because it is seen to
undergo Type I X-ray bursts \citep{tennantetal86,linaresetal10},
which occur in many accreting low magnetic field NSs in a
low-mass X-ray binary.
In contrast to \mcsnr, the spin period of Circinus~X-1 is not known,
and its highly variable X-ray luminosity can exceed
$10^{38}\mbox{ erg s$^{-1}$}$ \citep{linaresetal10,heinzetal15}.
A scenario in which Circinus~X-1 has $B\sim 10^{13}\mbox{ G}$
and $\dot{M}\sim 10^{-8}\,\Msun\mbox{ yr$^{-1}$}$ and
was born somewhat below its spin equilibrium period of $1\mbox{ s}$
would imply the NS started in the propeller phase, with a
timescale of $\sim 1000\mbox{ yr}$, and is now entering its
accretor/spin equilibrium phase.
For lower long-term accretion rates, a somewhat stronger
magnetic field would still yield the same result.

Finally one could consider applying the model described here to
low-mass X-ray binary (LMXB) systems.
LMXBs have much lower magnetic fields ($B\sim 10^8-10^9\mbox{ G}$),
which yield long ejector and propeller timescales,
$\tej\sim 10^8\mbox{ yr}$ and $\tprop\sim 10^9\mbox{ yr}$,
respectively.
However, since LMXBs are likely to be very old, it may be unlikely
for us to observe them early enough in their accretion history to
see potential effects of ejector and early propeller phases.

\section*{Acknowledgements}
The authors thank the anonymous referee for comments which led to
improvements in the manuscript.
WCGH and NA acknowledge support through grant ST/R00045X/1 from
the Science and Technology Facilities Council in the United Kingdom.
TMT acknowledges an AIAS--COFUND Senior Fellowship funded by the
European Union Horizon~2020 Research and Innovation Programme
(grant agreement no.~754513) and Aarhus University Research Foundation.



\bibliographystyle{mnras}
\bibliography{hmxbprop}


\bsp	
\label{lastpage}
\end{document}